\documentclass[a4paper,11pt]{article}
\pdfoutput=1 % if your are submitting a pdflatex (i.e. if you have
\usepackage{jcappub} 

\usepackage[T1]{fontenc} % if needed

\title{\boldmath GravitoMagnetic Field in Tensor-Vector-Scalar Theory}

\author{Qasem Exirifard}

\affiliation{ The Abdus Salam International Centre for Theoretical Physics (ICTP), Strada Costiera, Trieste, Italy
}

\emailAdd{exirifard@gmail.com}

\abstract{We  study the gravitomagnetism in the TeVeS theory. We  compute the gravitomagnetic field that   a slow moving mass distribution produces in its Newtonian regime. We report that the consistency between the TeVeS gravitomagnetic field and that predicted by the Einstein-Hilbert theory leads to a relation between the vector and scalar coupling constants of the theory. We observe  that requiring consistency between the near horizon geometry of a black hole in TeVeS and the image of the black hole taken Event Horizon Telescope leads to another relation between the coupling constants of the TeVeS theory and enable us to identify the coupling constants of the theory.}

\begin{document}
\baselineskip=0.77cm
\maketitle
\section{Introduction}
The missing mass problem in the galaxies can be solved either by the  modified  standard model of  the elementary particles (the dark matter paradigm),  or  the modified Newtonian dynamics/gravity (the dark force paradigm).  The later approach started by the Milgrom's theory of the Modified Newtonian Dynamics (MOND)\cite{MOND}, changed to the a-quadratic Lagrangian model for  gravity \cite{AQUAL} (AQUAL).  A generally covariant realization of the AQUAL model is the TeVeS theory  \cite{Bekenstein:2004ne}. The dynamical degrees of TeVeS are a scalar, a vector and a tensor. 

TeVeS equations of motion are more evolved compared to those of the Einstein-Hilbert theory. Its spherically static solution \cite{gr-qc/0502122} is known but its  exact Kerr-like solution is not yet known. In one hand, we do not know the exact solution for the gravito-magnetic field around the Earth in the TeVeS theory. In the other hand, however, we have acquired very precise empirical knowledge on the gravitomagnetic field.  In this paper, we  calculate the gravitomagnetic field of  a slow moving mass distribution  in the Newtonian regime of the TeVeS theory. We report that the consistency between this gravitomagnetic field and that predicted by the Einstein-Hilbert theory leads to a relation between the vector and scalar coupling constants.  We  translate the Lunar Laser Ranging measurement's data into a constraint on the deviation from this relation. 

The paper is provided as follows: Section  \eqref{TeVeSEQ} briefly reviews the field equations of the TeVeS theory. Section \eqref{TeVeSEarth} solves these equations at the linear order around the space-time  geometry of the Earth.  Section \eqref{Reports} reports that the consistency between this gravitomagnetic field and that predicted by the Einstein-Hilbert theory leads to a relation between the vector and scalar coupling constants of the theory. Last section before the conclusion's section reports how the image of the black hole constrains the theory.   

\section{Equations of motion of TeVeS}
\label{TeVeSEQ}
The dynamical degrees of freedom of  the TeVeS theory are a rank two tensor (geometrical metric) $g_{\mu\nu}$, a vector field $A_\mu$ and a scalar field $\phi$. Particles, however, move on the geodesic of the (Einstein) physical metric defined by
\begin{equation}\label{phymetric}
\tilde{g}_{ab} \,=\, e^{-2\phi} g_{ab} - 2 A_a A_b \sinh (2\phi)\,.
\end{equation}
The vector field is forced to be the unite timelike vector with respect to the Einstein metric
\begin{equation}
\label{Aunity}
\tilde{g}^{ab} A_a A_b = 1\,.
\end{equation}
We assume that matter is an ideal fluid. So the physical energy momentum tensor takes the form of 
\begin{equation}\label{T}
T_{ab} = \rho u_a u_b + p (\tilde{g}_{ab} + u_a u_b)\,,
\end{equation}
where $\rho$ is the proper energy, $p$ is the pressure and $u_a$ is the 4-velocity, all three expressed in the physical metric. In this work we set $p=0$ and use
\begin{eqnarray}
T_{00} & = & \rho\,, \\
T_{0i} & = & J_{i} = \rho v_i\,.
\end{eqnarray}
 The metric equation of motion reads
\begin{equation}\label{EG}
G_{ab} \, = \, 8 \pi G (T_{ab} + (1 - e^{-4\phi}) A^{\mu} T_{\mu(a} A_{b)} + \tau_{ab}) + \Theta_{ab}\,,
\end{equation}
where $G_{ab}$ is the Einstein tensor constructed out from the metric and its derivatives, round bracket represents symmetrization with respect to indices, e.g. $A_{(a}B_{b)} = A_a B_b + A_b B_a$. And $\tau_{ab}$ and $\Theta_{ab}$ stand for the contribution of the scalar and vector fields:
\begin{eqnarray}\label{tauab}
\tau_{ab} & \equiv & \sigma^2 \left( \phi_{,a} \phi_{,b} - \frac{1}{2} g^{\mu\nu} \phi_{,\mu}\phi_{,\nu} g_{ab}
- A^\mu \phi_{,\mu} (A_{(a} \phi_{,b)}- \frac{1}{2} A^\nu  \phi_{,\nu} g_{ab})  \right) - \frac{1}{4} G l^{-2} \sigma^4 \, F(k G \sigma^2) g_{ab},\nonumber \\\\
\Theta_{ab} & \equiv & K \left(
g^{\mu\nu} F_{\mu a} F_{\nu b} - \frac{1}{4}  F_{\mu\nu} F^{\mu\nu} g_{ab} 
\right)
- \lambda  A_{a} A_b\,,
\label{Thetaab}
\end{eqnarray}
where $F_{ab}$ represents the field strength of the vector field 
\begin{equation}
F_{ab} = \partial_a A_b - \partial_b A_a\,,
\end{equation}
and $\lambda$ is a Lagrange multiplier. Notice that  $K$ and $k$ are respectively the vector and scalar couplings, $l$ is a parameter and $F$ stands for the TeVeS function.  It is useful to define a function $\mu(y)$ by
\begin{equation}
-\mu F(\mu) -\frac{1}{2} \mu^2 F'(\mu) \,=\,y\,,
\end{equation}
using which expresses the equations of the scalar fields to
\begin{eqnarray}
  k G \sigma^2  & = & \mu (k l^2 h^{\mu\nu} \phi_{,\mu} \phi_{,\nu})\,, \\
  \label{Ephi}
\nabla_\beta \left(\mu (k l^2 h^{\mu\nu} \phi_{,\mu} \phi_{,\nu}) ~ h^{\alpha\beta} \,\phi_{,\alpha}\right) & =& k G [g^{\mu\nu} + (1 + e^{-4\phi}) A^\mu A^\nu] T_{\mu\nu}\,,
\end{eqnarray}
where 
\begin{equation}
h^{ab} \equiv  g^{ab} - A^a A^b \,.
\end{equation}
The equation of motion for the vector field reads
\begin{equation}\label{EV}
K \nabla_b F^{a b} + \lambda A^a + 8 \pi G \sigma^2 A^b \nabla_b\phi~ \nabla^a \phi \,=\, 8\pi G (1- e^{-4\phi}) A_b T^{ab}\,.
\end{equation}
The equation of motion for the Lagrange multiplier is derived by contracting \eqref{EV} with $A_a$ and utilizing \eqref{Aunity}:
\begin{equation}\label{Elambda}
 \lambda \,=\, 8\pi G (1- e^{-4\phi}) A_a A_b T^{ab}-K A_a \nabla_b F^{a b} - 8 \pi G \sigma^2 (A^b \nabla_b\phi)^2\,.
\end{equation}

\section{TeVeS theory in the space-time geometry around the Earth}
\label{TeVeSEarth}
We shall consider the Newtonian regime of TeVeS theory wherein $\mu\approx 1$. We also assume that $l$ is large enough to ignore  the last term in the right hand side of \eqref{tauab}. In other words we solve the TeVeS equations of motion in the Newtonian regime of the theory. Thus we set:
\begin{eqnarray}
\label{tauabl}
\tau_{ab} & = & \sigma^2 \left( \phi_{,a} \phi_{,b} - \frac{1}{2} g^{\mu\nu} \phi_{,\mu}\phi_{,\nu} g_{ab}
- A^\mu \phi_{,\mu} (A_{(a} \phi_{,b)}- \frac{1}{2} A^\nu  \phi_{,\nu} g_{ab})  \right)+ O(\frac{1}{l})\,,\\
\sigma^2 \, &=& \,\frac{1}{k G}+ O(\frac{1}{l}) \,.
\end{eqnarray}
Note that $O(\frac{1}{l})=O(\frac{c^2}{l}) = O(\frac{a_0}{a_N})$ where $a_0$ is the critical acceleration of the MOND theory and $a_N$ is the Newtonian gravitational field strength.  
We are interested in the solution representing the space-time geometry around the Earth. To this aim we consider that in the absence of the Earth  holds :
\begin{subequations}
\label{cosmos}
\begin{eqnarray}
g_{ab} &=& \eta_{ab} \,,\\
\phi& =& 0  \,,\\
A_\mu &=& \delta_\mu^0 \, ,\\
\lambda &=& 0\,, 
\end{eqnarray}
while the Energy-Momentum tensor \eqref{T} reads
\begin{equation}
p= \rho = 0\,. 
\end{equation}
\end{subequations}
Notice that \eqref{cosmos} solves  \eqref{EG}, \eqref{Ephi}, \eqref{EV} and \eqref{Elambda}. 
The Earth introduces a perturbation to \eqref{cosmos}: 
\begin{subequations}
\label{Earth}
\begin{eqnarray}
g_{ab} &=& \eta_{ab} + \epsilon h_{ab}^{(1)} + O(\epsilon^2) \,,\\
\phi &=& 0 + \epsilon \phi^{(1)} + O(\epsilon^2)  \,,\\
\label{Apert}
A_\mu &=& \delta_\mu^0 + \epsilon A_\mu^{(1)}+ O(\epsilon^2) \\
\lambda &=& 0 + \epsilon \lambda^{(1)}+ O(\epsilon^2)\,, 
\end{eqnarray}
and
\begin{eqnarray}
p= 0+ \epsilon p^{(1)}\,, \\
\rho = 0 + \epsilon \rho^{(1)} \,,
\end{eqnarray}
\end{subequations}
where $\epsilon$ is the systematic parameter of the perturbation. In the approximation we are working in $p^{(1)}=0$. 
Inserting the perturbative series in the Einstein metric yields:
\begin{equation}\label{phymetric1}
\tilde{g}_{ab} \,=\, \eta_{ab}+ \epsilon (h^{(1)}_{ab} -2  \phi^{(1)} \eta_{ab} - 4 \delta^0_a \delta^0_b \phi^{(1)})+O(\epsilon^2)\,.
\end{equation}
The inverse of the Einstein metric holds
\begin{equation}\label{phymetric2}
\tilde{g}^{ab} \,=\, \eta^{ab}- \epsilon (\eta^{a\alpha}\eta^{b\beta}h^{(1)}_{\alpha\beta} -2  \phi^{(1)} \eta^{ab} - 4 \eta^{a\alpha} \eta^{b\beta} \delta^0_\alpha \delta^0_\beta \phi^{(1)})+O(\epsilon^2)\,.
\end{equation}
Inserting \eqref{phymetric2} and  \eqref{Apert} into \eqref{Aunity} gives an equation for $A_0^{(1)} $ which is solved by
\begin{equation}\label{22}
A_0^{(1)} = \frac{1}{2} h^{(1)}_{00} -  \phi^{(1)}\,.
\end{equation}
Inserting the perturbative series in \eqref{Elambda} and keeping the first order perturbation gives
\begin{equation}
\label{lambda0Eq}
\lambda^{(1)} = - K \nabla_b F^{(1)0b}\,.
\end{equation}
The space-time geometry around the Earth is stationary at the leading approximation:
\begin{subequations}
\label{stationary}
\begin{eqnarray}
A^{(1)}_\mu &\equiv &A^{(1)}_\mu(\vec{x})\,,\\
h^{(1)}_{\mu\nu}&\equiv &h^{(1)}_{\mu\nu}(\vec{x})\,,\\
\phi^{(1)}&\equiv &\phi^{(1)}(\vec{x})\,.
\end{eqnarray} 
\end{subequations}
Utilizing \eqref{stationary} in  \eqref{lambda0Eq} yields:
\begin{equation}
\label{25}
\lambda^{(1)} = - K \nabla^2 A_0^{(1)}\,,
\end{equation}
where in the Cartesian coordinates 
\begin{equation}  
\nabla^2 = \sum_{i=1}^3(\partial_i)^2\,.
\end{equation}
Inserting \eqref{Earth} into \eqref{tauabl} and \eqref{Thetaab} yields:
\begin{eqnarray}
\tau_{ab} &=& 0+ O(\epsilon^2,\frac{1}{l})\,, \\
\Theta_{ab} & = & 0 - \epsilon \lambda^{(1)} \delta_{a}^{0} \delta_b^0 + O(\epsilon^2, \frac{1}{l})\,,
\end{eqnarray}
inserting which into \eqref{EG} results
\begin{equation}
\label{G=T}
G_{ab}^{(1)} = 8 \pi G T_{ab}^{(1)} - \lambda^{(1)} \delta_a^0 \delta_b^0\,.
\end{equation}
where $G_{ab}^{(1)}$ represents the linearized Einstein tensor constructed out from $h_{\mu\nu}$. 
Now let the gravitoelectric and gravitomagnetic fields be defined as usual: 
\begin{eqnarray}\label{30}
\Phi &=& \frac{1}{2} h_{00}\,,\\
A_i &=& h_{0i}\,.\label{30A}
\end{eqnarray}
Then  in the GravitoElectroMagnetism's approximation eq. \eqref{G=T} is re-expressed to
\begin{eqnarray}
\label{32}
\nabla^2 \Phi &=& 4 \pi G \rho -\frac{1}{2} \lambda^{(1)}\,, \\
\label{33}
\nabla^2 A &=& \frac{16 \pi G}{ c^2} \vec{j} \,,
\end{eqnarray}
where $\vec{j}$ is the current density: $\vec{j}=\rho \vec{v}$. Inserting the perturbative series of \eqref{Earth} in \eqref{Ephi} results
\begin{equation}\label{34}
\nabla^2 \phi^{(1)} = k G \rho\,.
\end{equation}
Eq \eqref{22}, \eqref{25},\eqref{32} and \eqref{34} imply that 
 \begin{equation}
 \label{35}
 (1- \frac{K}{2}) \nabla^2 \Phi + \frac{K}{2} \nabla^2 \phi^{(1)} =  4\pi G \rho\,.
 \end{equation}
Using \eqref{34} in \eqref{35} gives:
 \begin{equation}
 \label{36}
 \nabla^2 \Phi =  4\pi G  \frac{1-\frac{k K}{8 \pi}}{1- \frac{K}{2}}~\rho\,.
 \end{equation} 
 Recalling that  the Newtonian potential $\Phi_N$ solves
\begin{equation}
\nabla^2 \Phi_N = 4 \pi G \rho\,,
\end{equation}
where $G$ is the Newton's constant, $\Phi$ and $\phi^{(1)}$ can be expressed in term of $\Phi_N$
\begin{eqnarray}\label{3.23last}
\Phi & =& \frac{1-\frac{k K}{8\pi}}{1- \frac{K}{2}} \Phi_N \,,\\
\phi^{(1)} & = & \frac{k}{4\pi} \Phi_N\,.\label{3.24last}
\end{eqnarray}
Notice that particles move on the geodesic of the Einstein (physical) metric \eqref{phymetric1}. The physical gravitoelectric and gravitomagnetic field, therefore, must be defined by the physical metric:
\begin{eqnarray}
\Phi_{\mbox{\tiny Phy}} &\equiv& \frac{1}{2} \tilde{g}_{00}^{(1)}\,,\\
A_i^{\mbox{\tiny Phy}} &\equiv& \tilde{g}_{0i}\,.
\end{eqnarray} 
Eq. \eqref{phymetric1}, \eqref{30}, \eqref{30A}, \eqref{3.23last} and \eqref{3.24last} then imply  that
\begin{eqnarray}
\Phi_{\mbox{\tiny Phy}} &=& \Phi - \phi^{(1)} = \left(\frac{1-\frac{k K}{8\pi}}{1-\frac{K}{2}}-\frac{k}{4\pi}\right) \Phi_N\,,\\
A_i^{\mbox{\tiny Phy}} &= & A_i\,.
\end{eqnarray}
The physical quantities, therefore, solve:
\begin{subequations}
\label{44}
\begin{eqnarray}
\label{44a}
\nabla^2 \Phi_{\mbox{\tiny Phy}}  &=& \left(\frac{1-\frac{k K}{8\pi}}{1-\frac{K}{2}}-\frac{k}{4\pi}\right)  4\pi G \rho\,, \\
\nabla^2 A_i^{\mbox{\tiny Phy}} &= &  \frac{16 \pi G}{ c^2} \vec{j}\,.
\end{eqnarray}
\end{subequations}
We observe that for $k=4\pi$, the physical gravitoelectric field vanishes. In the limit of the $K\to 2$, the physical gravitomagnetic field diverges.%\footnote{It is tempting to suggest that the limit of $K\to 2$ of the TeVeS theory may provide a framework accommodating  some of the quantum gravity's features.}  
We assume that $k\neq 4\pi$ and $K\neq 2$. In these regimes, the $\frac{1}{r^2}$behavior of the $\Phi_{\mbox{\tiny Phy}}$ is used  to measure the Newton's constant.  Because of  \eqref{44a} the measured value of the Newton's constant in the TeVeS theory  reads:
\begin{equation}
G^{\mbox{\tiny Obs.}} \equiv \left(\frac{1-\frac{k K}{8\pi}}{1-\frac{K}{2}}-\frac{k}{4\pi}\right) G  \,.
\end{equation}
Rewriting \eqref{44} in term of the observed value of the Newton's constant yields
\begin{subequations}
\label{46}
\begin{eqnarray}
\nabla^2 \Phi_{\mbox{\tiny Phy}}  &=&   4\pi G^{\mbox{\tiny Obs.}} \rho\,, \\
\nabla^2 A_i^{\mbox{\tiny Phy}} &= & (1+\beta) \frac{16 \pi G^{\mbox{\tiny Obs.}}}{ c^2} \vec{j}\,,
\label{46b}
\end{eqnarray}
\end{subequations}
where
\begin{equation}\label{betadefine}
\beta \equiv \left(\frac{1-\frac{k K}{8\pi}}{1-\frac{K}{2}}-\frac{k}{4\pi}\right)^{-1}-1
\end{equation}
The Einstein-Hilbert gravity, however, predicts that 
\begin{subequations}
\begin{eqnarray}
\nabla^2 \Phi_{\mbox{\tiny EH}}  &=&   4\pi G^{\mbox{\tiny Obs.}} \rho\,, \\
\label{47b}
\nabla^2 A_i^{\mbox{\tiny EH}} &= & \frac{16 \pi G^{\mbox{\tiny Obs.}}}{ c^2} \vec{j}\,.
\end{eqnarray}
\end{subequations}
Comparing \eqref{46b} and \eqref{47b} proves that the TeVeS theory and the Einstein-Hilbert gravity generally predict different values for the gravitomagnetic field around the Earth.  The deviation is measured in term of the $\beta$ parameter  defined in \eqref{betadefine}.

\section{Consistency between  Einstein-Hilbert gravity and TeVeS theory}
\label{Reports}
The theoretical consistency between \eqref{46b} and \eqref{47b} requires that 
\begin{equation}\label{beta=0}
\beta = 0.
\end{equation}
 In other words the gravitoelectric field of the Newtonian regime of TeVeS theory coincides to that predicted by Einstein-Hilbert gravity only for \eqref{beta=0}.  Using \eqref{betadefine} in  \eqref{beta=0} results
 \begin{equation}
 \frac{4 \pi - 2 \pi K}{4 \pi- k} = 1\,,
 \end{equation}
 which for $k\neq 4\pi$ and $K\neq 2$ is solved by 
 \begin{equation}\label{Kk}
 K = \frac{k}{2\pi }\,.
 \end{equation}

\section{The Shadow of a black hole in TeVeS theory}
The  Event Horizon Telescope  \cite{Akiyama:2019cqa} has recently announced  that the radius of the event horizon of the M87  black hole  is compatible with that of the Schwarzschild metric.  This suggests that in any theory of gravity, the near horizon geometry of a stationary black hole of a given mass must coincide to that of the Schwarzschild metric with the same mass.  Einstein-Hilbert theory meets this criterion since the Schwarzschild metric is its exact solution. It, however, provides a non-trivial constraint  on the parameters of a general modified theory of gravity. J. W. Moffat and V. T. Toth   have shown that the consistency of the near horizon geometry of a black hole in MOG and the Schwarzschild metric specifies the value of the $\alpha$ parameter of the MOG theory \cite{Moffat:2019uxp}. We would like to  consider the TeVeS theory \cite{Bekenstein:2004ne} and study  the compatibility of its  near horizon geometry with  the Schwarzschild metric.  

We observe that the non-rotating static solution of the TeVeS in its strong gravity regime, for the branch of the solution that  leads to the observed values of the $\beta$ and $\gamma$ PPN coefficients, can be made identical to that of the Einstein gravity, as reported in the section V of \cite{gr-qc/0502122}.   We first should impose regularity on the physical event horizon:
\begin{equation}\label{eq2}
\frac{r_c}{r_g} \,=\, \frac{1}{4} +\frac{\kappa}{8\pi}\frac{ G\,m_s}{ r_g}\,, ~~~(\text{Eq. (78) of \cite{gr-qc/0502122}})
\end{equation}
where $r_c$ is the Schwarzschild radius 
\begin{eqnarray}\label{eq3}
16 (\frac{r_c}{r_g})^2 = 1 + \frac{\kappa}{\pi}(\frac{G m_s}{r_g})^2 - \frac{K}{2}\,,~~~(\text{Eq. (71) of \cite{gr-qc/0502122}})\,,
\end{eqnarray}
where $Gm_s$ encodes the fall off of the scalar field.  $r_g$ encodes the mass of the black hole in the TeVeS theory.
We next require it to be the same as the Schwarzschild radius:
\begin{equation}\label{rcmg}
r_g = r_c\,.
\end{equation}
In so doing the near horizon geometry of a black hole for a given mass in TeVeS theory coincides to that of the Einstein theory. Solving  (\ref{eq2}),(\ref{eq3}) and (\ref{rcmg}) gives:
\begin{equation}\label{BHK}
K \,=\, -30 + \frac{72 \pi}{\kappa}\,.
\end{equation}
We notice that the consistency between the TeVeS gravitomagnetic field and that predicted by the Einstein-Hilbert theory and measured by the Gravity Probe B \cite{Everitt:2011hp} demands: 
\begin{equation}\label{gravomagnetic}
K \,=\, \frac{\kappa}{2\pi}\,.
\end{equation}
Eq. (\ref{gravomagnetic}) and (\ref{BHK}) then enable us to find the values of the coupling constants of the theory: 
\begin{eqnarray}
K&=& 3 (\pm\sqrt{29}-5) \,,\\
\kappa&=& 6 \pi (\pm\sqrt{29}-5) \,,
\end{eqnarray}
inserting which into (\ref{eq2}) gives
\begin{equation}
\frac{Gm_s}{r_g} \,=\, \frac{1}{4} (5\pm\sqrt{29})\,,
\end{equation}
One may wish not to consider a negative value for $m_s$ and this uniquely identifies the coupling constants. This is the first time that the coupling constants of the TeVeS theory are fixed. We observe that the values of $K,\kappa$ are large while the literature most often assumed that they are small.   We further  notice they satisfy
\begin{equation}
K < \frac{2\kappa}{\pi} (\frac{G m_s}{r_g})^2 ~~~(\text{Eq. (73) of \cite{gr-qc/0502122}})\,,
\end{equation} 
so the total energy density of vacuum contributed by the scalar and vector fields remains positive in the whole of space-time. This is very nontrivial and points that TeVeS can be both consistent with the data and be free of quantum instability. 

\section{Conclusion}
We have calculated the gravitomagnetic field around a slow moving mass distribution in the Newtonian regime of the TeVeS theory. We have shown that requiring consistency between the gravitomagnetisem in the Einstein-Hilbert theory and the TeVeS theory leads to a relation between the scalar and vector coupling constants of the TeVeS theory. We have observed  that requiring consistency between the near horizon geometry of a black hole in TeVeS and with that reported by the Event Horizon Telescope leads to another relation between the coupling constants of the TeVeS theory. We have identified  the coupling constants of the TeVeS theory.

 \section*{Acknowledgements}
  I would like to thank Paolo Creminelli and Viktor T. Toth  for discussions and email correspondences.   I would like to thank ICTP for its nice hospitality.

 \providecommand{\href}[2]{#2}\begingroup\raggedright
\end{document}